\documentclass[conference]{IEEEtran}

\usepackage{multirow}
\usepackage[font=scriptsize,caption=false,labelsep=space]{subfig}
\usepackage{mathrsfs}
\usepackage{graphicx,float,wrapfig,epstopdf,amsmath}
\usepackage[]{algorithmicx}
\usepackage{algpseudocode,algorithm}
\usepackage{balance}
\usepackage{mathtools,lipsum}
\usepackage{amsmath}
\usepackage{amssymb}
\usepackage{amsthm}
\usepackage{graphicx}
\usepackage{epstopdf}
\usepackage{times}
\usepackage{textcomp,cite}

\usepackage{url}
\usepackage{hyperref}

\usepackage{multirow}
\usepackage{threeparttable}
\graphicspath{{./figures/}}

\usepackage[table]{xcolor}
\definecolor{intnull}{RGB}{213,229,255}
\definecolor{inteins}{RGB}{128,179,255}
\definecolor{color1}{RGB}{199,209,232}
\definecolor{color2}{RGB}{230,231,233}

\begin{document}

		\title{Vehicular Networks for Combating A Worldwide Pandemic: Preventing the Spread of COVID-19}

		\author{\IEEEauthorblockA{Ahmet M. Elbir$^{1,2}$,
			Gokhan Gurbilek$^{3,4}$,
			Burak Soner$^{3,4}$,
			Anastasios K. Papazafeiropoulos$^{5,2}$,\\
			Pandelis Kourtessis$^{5}$,
			Sinem Coleri$^{3}$}
		\IEEEauthorblockA{
			$^1$: Electrical and Electronics Engineering, Duzce University, Duzce, Turkey\\
			$^2$: University of Luxembourg, Luxembourg \\
			$^3$: Electrical and Electronics Engineering,  	Koc University, Istanbul, Turkey\\
			$^4$: Koc University Ford Otosan Automotive Technologies Laboratory (KUFOTAL), Istanbul, Turkey\\
			$^5$: University of Hertfordshire, Hatfield, U. K.
		}
		}

				\maketitle
			\begin{abstract}
			As a worldwide pandemic, the coronavirus disease-19 (COVID-19) has caused serious restrictions in people's social life, along with the loss of lives, the collapse of economies and the disruption of humanitarian aids. Despite the advance of technological developments, we, as researchers, have witnessed that several issues need further investigation for a better response to a pandemic outbreak. Therefore, researchers recently started developing ideas to stop or at least reduce the spread of the pandemic. \textcolor{black}{While there have been some prior works on wireless networks for combating a pandemic scenario, vehicular networks and their potential bottlenecks have not yet been fully examined. Furthermore, the vehicular scenarios can be identified as the locations, where the social distancing is mostly violated. With this motivation, this article provides an extensive discussion on vehicular networking for combating a pandemic.} We provide the major applications of vehicular networking for combating COVID-19 in public transportation, in-vehicle diagnosis, border patrol and social distance monitoring. Next, we identify the unique characteristics of the collected data in terms of privacy, flexibility and coverage, then highlight corresponding future directions in privacy preservation, resource allocation, data caching and data routing. We believe that this work paves the way for the development of new products and algorithms that can facilitate the social life and help controlling the spread of the pandemic.
		\end{abstract}

\section{Introduction}
The coronavirus disease-19 (COVID-19) has spread globally, causing millions of confirmed cases and  deaths  according to the world health organization (WHO). \textcolor{black}{The pandemic has caused serious difficulties in people's social life, such as unprecedented numbers in unemployment, travel restrictions and strict lockdowns~\cite{Verma2020May}.} In order to stop or at least slow down the spread of the COVID-19, authorities initiated precautions to limit the social/physical interactions by social distancing, i.e., keeping the distance among people in social space at a certain level, specifically $1$ meter ($3$ ft). Social distancing helps to reduce the effect of spreading the disease  from infected to healthy person. The need for monitoring the social interaction of people  has motivated researchers to develop social network aware wireless network architectures~\cite{detectingMobilityCOVID,social_IoT}. Authors in \cite{detectingMobilityCOVID} predict the \emph{at-risk} regions regarding the spread of COVID-19 by utilizing the mobility information of the mobile users in cellular networks. In \cite{social_IoT}, the proximity sensor data collected from the individuals in a crowded area are used to monitor and predict the spread of COVID-19. The extension of these social wireless network architectures for the support of vehicular networking needs further investigation to meet the requirements of vehicle based COVID-19 applications while incorporating high mobility and intermittent connectivity. 

Vehicular networking needs to be integrated into the communication architecture built for  combating a pandemic to trace the spread of disease, detect the violation of social distancing and enable health care applications, since most people use public and private transportation for their daily life. The WHO identifies three basic steps for the tracing of the disease in social life, which are contact identification, contact listing, and follow-up. Vehicular networks can be used for this purpose since most people use public and private transportation, where the interaction of the infection is almost inevitable~\cite{BibEntry2021Mar,Wirth2020Nov}. The WHO further identifies vehicles as the locations, where people violate the social distancing most. Using the sensors on board in a vehicle, reliable data regarding the social distancing can be collected from the people, and then shared with the authorities. 	Furthermore, vehicular communication is  required for health care applications in case of a pandemic outbreak for the path planning of emergency vehicles and the transmission of the vitals of the patients from the emergency vehicle in the field to the hospital. Unmanned aerial vehicles (UAV) can also be used in social distance monitoring~\cite{uavNews1,Chamola2020May} and be integrated with vehicular communication for more accurate predictions.

\begin{table*}[h]
\caption{Summary of Major Applications of Vehicular Communications for Preventing the Spread of A Pandemic}
\label{tableApps}
\centering
\begin{tabular}{p{0.21\textwidth} p{0.21\textwidth} p{0.21\textwidth}  p{0.21\textwidth} }
	\hline
	
	\hline
	\centering Public Transportation \textcolor{black}{(U2V)} & \centering In-Vehicle Diagnosis \textcolor{black}{(U2V)} & \centering Border Patrol \textcolor{black}{(V2I)}&        UAV-assisted Social Distance Monitoring \textcolor{black}{(V2I)}\\
	\hline
	\hline
	The health status, contact information of the passengers and proximity sensor readings can be collected in public transportation then transmitted to a data center for further analysis to monitor the spread of the pandemic and make predictions.   &
	
	In medical vehicles, infection test and diagnosis can be made on patients without necessarily accommodating them in the hospitals where the spread of the disease is a risk. The collected diagnosis data in the vehicle can then be shared with medical authorities. &
	
	At the checkpoints of cities, vehicles can collect the data from the passengers and share it with road side infrastructures automatically, mitigating the long queues of vehicles and human interactions.  &
	
	UAVs can be used to monitor social distancing together with vehicles on the streets to benefit from the multiple sensors in the vehicles, such as cameras, proximity sensors and radars/lidars. Additionally, the collected data can be used for the prediction of the spread of the disease.   \\
	\hline
\end{tabular}
\end{table*}

In the aforementioned scenarios, vehicular networks can be used to collect and process the data by exploiting various communication pairs with vehicles in the heterogeneous architecture~\cite{multihopTVT}. User-to-vehicle (U2V) communication can be used to collect data from the passengers in a vehicle. Vehicle-to-infrastructure (V2I) communication can be used to convey  information from the vehicles to the data centers via road-side units (RSUs) or cellular base stations. Furthermore, vehicle-to-vehicle (V2V) communication can be used to disseminate the information to the other vehicles for the applications like emergency warning and route planning. Employing  vehicular communication  can accelerate the process of detecting and monitoring the infection while reducing human interactions, preventing the spread of the disease. \textcolor{black}{Hence, equipped with several sensors,  reliable data about the health status of people in the vehicles can be collected and  used for monitoring and controlling the spread of the pandemic. As a result, vehicular communication can bring benefits over the use conventional wireless networking, thereby, collecting more accurate and larger diversity of data. In addition to such advantages, the usage of vehicular applications in a pandemic also brings difficulties and challenges, such as privacy concerns and new communication protocols. With this motivation, in this article, we investigate the usage of vehicular communications in combating a pandemic, present various vehicular	networking applications in different communication	environments, and identify related research challenges. While there are other works on the general use of wireless technologies for combating social distancing, such as  WiFi, cellular, Bluetooth, ZigBee and RFID~\cite{covid_survey_Access,covid_survey_Access2}. None of these studies consider the additional challenges of vehicular networking, including high network dynamics, intermittent connectivity and heterogeneous architecture in vehicular communication networks including U2V, V2I and V2V.}

This article focuses on vehicular networks for preventing the spread of a highly contagious disease such as COVID-19. In the remaining of this article, Section II provides a survey of prior works related to the usage of general wireless networks for COVID-19 and the usage of vehicular networking for general health applications. Section III provides vehicular networking applications and their requirements for combating a pandemic, which can take place in public transportation, in-vehicle diagnosis, border patrol and social distance monitoring. Section IV identifies the related design and research challenges in privacy preserving mechanisms, resource allocation, data caching, vehicular data routing and machine learning (ML) solutions, and provide future research directions. Finally, Section V summarizes the article with conclusions.

\section{Prior Art}
In the following, we discuss the prior works in two categories, namely, wireless networks for COVID-19 and vehicular networks for health applications.

\subsection{Wireless Networks for COVID-19} 

Due to its recent emerge in 2019, there have been a limited number of works on the usage of wireless networks for  COVID-19~\cite{detectingMobilityCOVID,social_IoT,covid_survey_Access}. In~\cite{detectingMobilityCOVID}, authors exploit the ultra-dense small-cell deployment architecture of heterogeneous networks (HetNets) to determine the \emph{at-risk} regions of a geographical area by inferring the crowdedness and the mobility information of smart phone users in the cells. The key idea is that the cellular area is regarded as \emph{at-risk} if it involves strong mobility and high number of mobile users. Conversely, the sparse areas, with people remaining at home, are regarded as \emph{not at-risk}. A social internet of things (SIoT) approach is proposed to determine the contact intensity of people from the proximity sensors  in~\cite{social_IoT}. The sensor dataset, collected in a school and a museum (not specific to COVID-19), includes the contact intensity of the individuals and the number of infected cases for different closure time intervals. The collected data are modeled as a weighted undirectional graph (WUG), and then fed to an ML model yielding the set of \emph{at-risk} individuals at the output. In addition, \cite{covid_survey_Access}  surveys the wireless technologies for combating social distancing, such as  WiFi, cellular, Bluetooth, ZigBee and RFID, without specific interest on the vehicular networks. 

The extension of this approach to include vehicles requires the identification of the mobility pattern of vehicles, the type of vehicles indicating whether they are public or private, the location of the users within the vehicles, and incorporating the communication and processing of these data by the integration of the vehicular communication architecture, including U2V, V2I and V2V, into the heterogeneous architecture.



\subsection{Vehicular Networking for Health Applications}

The vitals of the patients need to be transmitted from the emergency vehicles to the hospitals for faster diagnosis and treatment. For this reason, vehicular networks have been examined in various prior works in the context of health applications for the  transmission of health data~\cite{vehiHealth,vehicular_bodynetwork2}. In~\cite{vehiHealth},  the vehicular communication between the ambulance and the hospital is studied. An emergency routing protocol is proposed  to transfer the patient current status information to the nearby hospital through V2V links with smaller link breakage probability and shorter path. In \cite{vehicular_bodynetwork2} together with wireless body sensor networks (WBSNs) in a smart ambulance application, the vitals of the patients are measured and collected by WBSN, and transmitted to the hospital from the ambulance via a V2I communication environment. However, none of these studies consider the additional requirements of a pandemic scenario in privacy, flexibility, coverage and prioritization. 

\begin{figure*}[t]
\centering
{\includegraphics[draft=false,width=\textwidth]{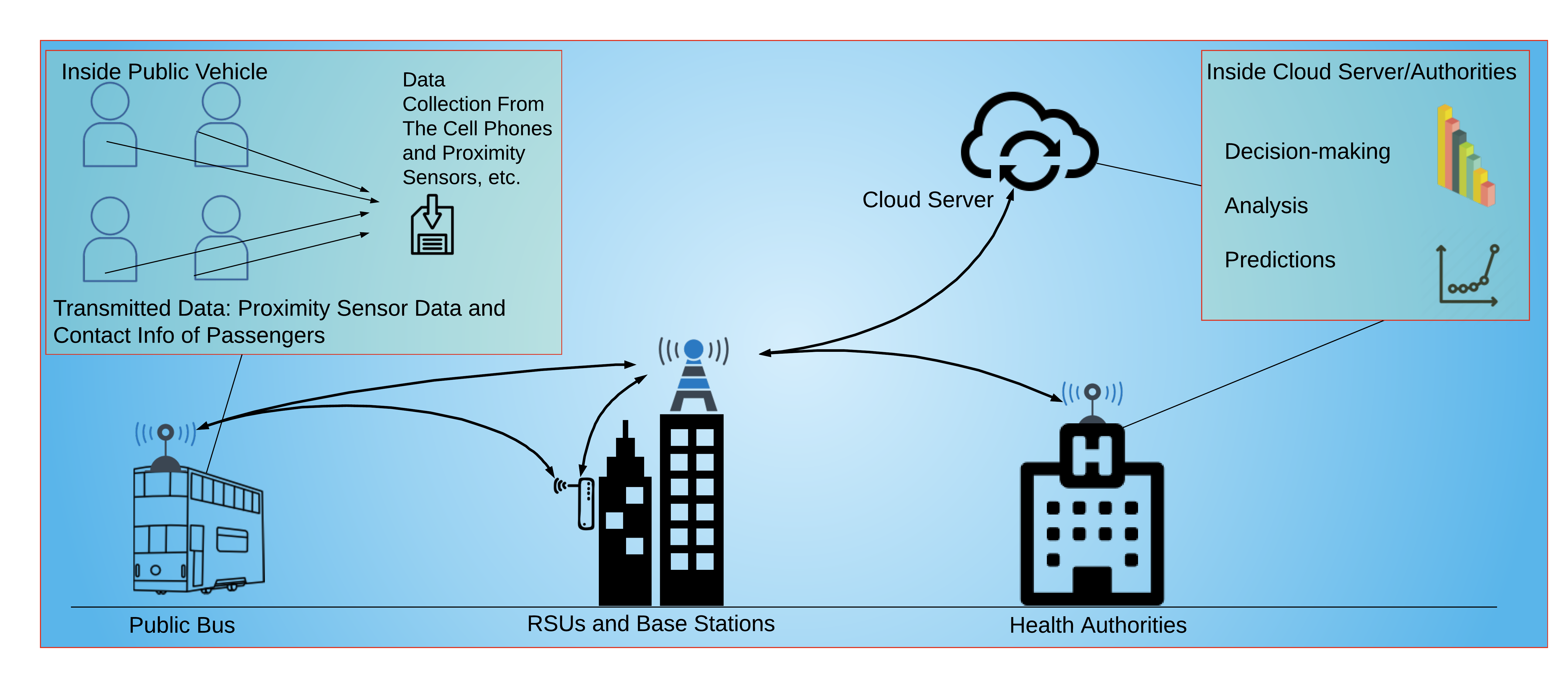} }
\caption{Infection tracing and data collection in public transportation. \textcolor{black}{The identification, travel history and health information of the passengers can be collected via U2V communication in  public transportation vehicles.}
}
\label{figPublicTransportation}
\end{figure*}

\section{Vehicular Networking Applications and Requirements for     Preventing Pandemic Spread}
In this section, we exploit the structure of vehicular networks specifically tailored for the prevention of a pandemic. First, we enlist the major vehicular applications and use cases that can provide a successful response to stop or at least reduce the spread of a pandemic, and then provide their data and network requirements.

\subsection{Applications and Use Cases}	
The major vehicular applications for preventing the pandemic spread are investigated under four categories: 1) Public transportation, 2) In-vehicle infection diagnosis, 3) Border patrol and 4) UAV-assisted crowd monitoring. \textcolor{black}{In Table~\ref{tableApps}, we summarize these categorized as well as indicating their application focus on either U2V (data collection from the individuals on vehicle) or V2I (collected data transmission from the vehicle to infrastructure).}

\begin{figure*}[h]
\centering
{\includegraphics[draft=false,width=\textwidth]{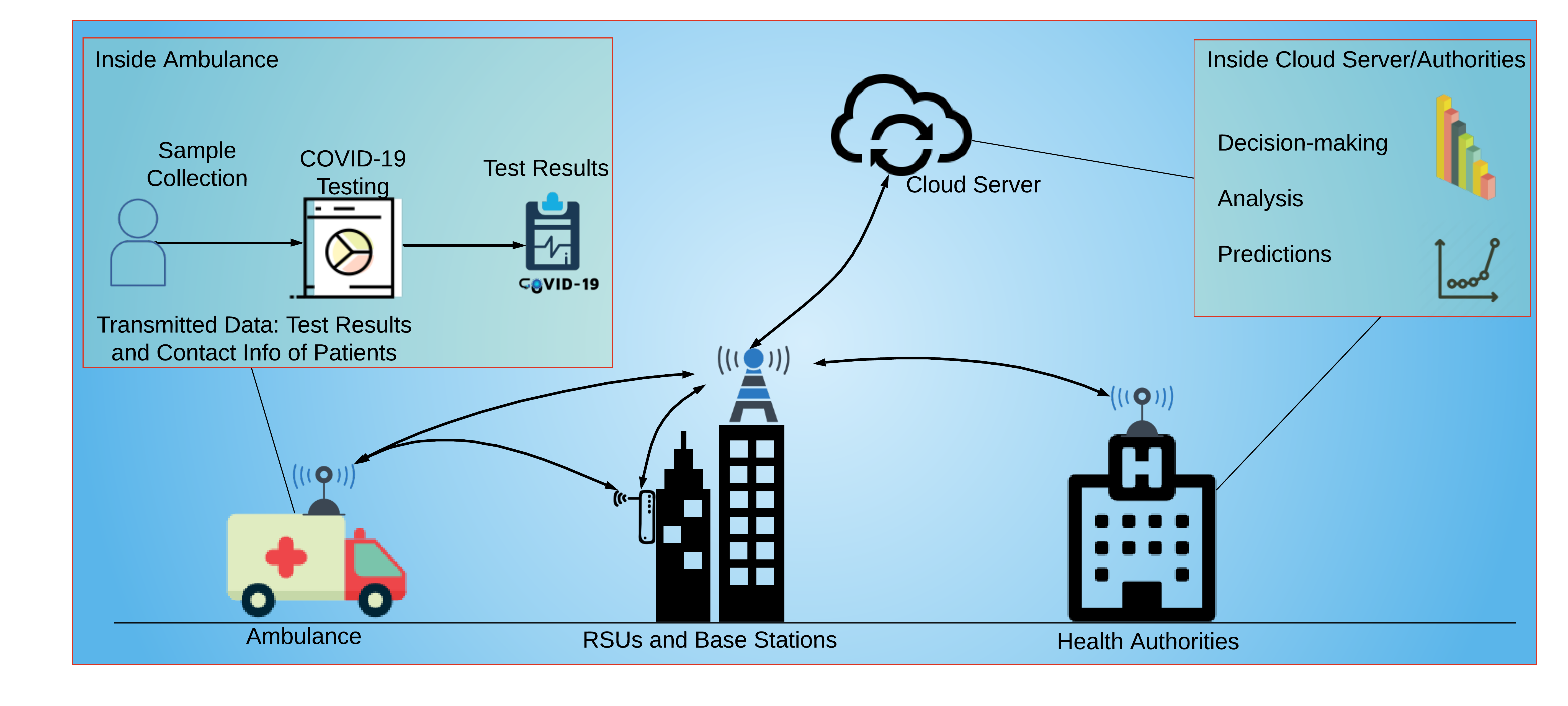} }
\caption{In-vehicle COVID-19 diagnosis. \textcolor{black}{The samples are collected via U2V communication from the patients and tested for COVID-19 within the vehicle without necessarily accommodating them in the hospitals.}
}
\label{figInVeh}
\end{figure*}

\subsubsection{Public Transportation}

In case of a pandemic, the tracing of the disease in social life is crucial in order to monitor and control the spread of the disease. In public transportation, the passenger data can be collected and then transmitted to a data server and  used by authorities, as illustrated in Figure~\ref{figPublicTransportation}. The collected data can include the identification, travel history, health information of the passengers, including the contact information, the status of having chronic illness, whether being infected by the disease, and proximity sensor readings identifying the distance among the passengers. The identification, travel history and health information of the passengers can be collected via U2V communication by either reading the identity card of the passengers or their mobile phones with a device located in  public transportation vehicles. On the other hand, the proximity sensors can be placed in various locations within the vehicle and send either raw data or processed data to the vehicle via U2V communication. Then, the data can be transmitted to the cloud servers to be further processed and stored by exploiting both V2V and V2I communication. These data are used by the government authorities to monitor the spread of the disease such that  if the infection is detected in an individual, then the others who have close relation, and have used the same public transportation are informed. This may be followed by  stay-at-home measures or even strict lockdowns depending on the spread of the disease and capacity of the hospitals. Moreover, an infection risk factor can be calculated using these data and the vehicle passenger capacity information, so that the public transportation vehicles above a certain risk factor can be prevented to board new passengers. Furthermore, individual infection risk factor and infection severity factor can be calculated for the passengers by using an artificial neural network (ANN) based data processing. For example, the collected data from the passengers can be used to train an ANN whose output indicates the risk factor of the vehicle in public transportation. 
\begin{figure*}[t]
\centering
{\includegraphics[draft=false,width=\textwidth]{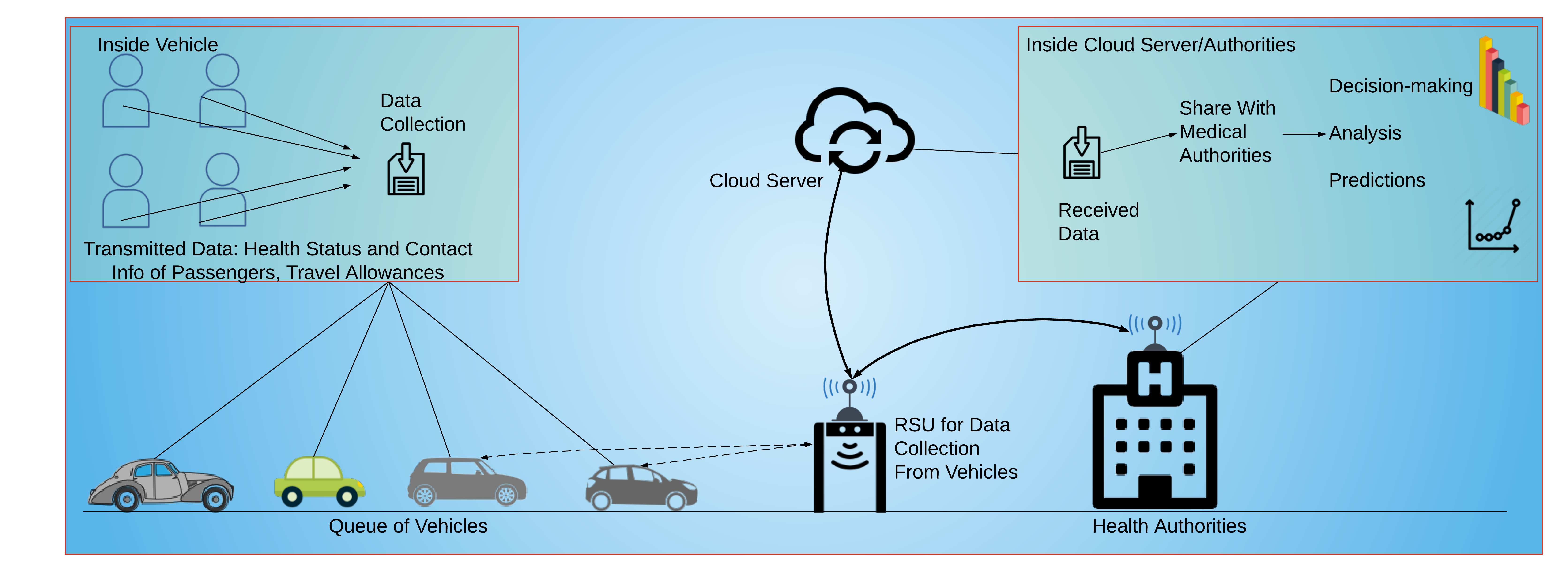} }
\caption{Border patrol at city checkpoints. \textcolor{black}{V2I networking can be used to automatically transfer the passenger data from the vehicles to the RSUs or police cars on the side of the road.}
}
\label{figBorderPatrol}
\end{figure*}


\subsubsection{In-Vehicle Infection Diagnosis}
We have witnessed that most of the hospitals operate at full capacity during a pandemic, hence, the diagnosis and treatment of the new-coming patients becomes a real challenge such that some of the governments have built temporary medical infrastructures. In this respect, the in-vehicle diagnosis of the disease can be helpful, as illustrated in Figure~\ref{figInVeh}. Specifically, samples are collected from the patients and tested for COVID-19 within the vehicle without necessarily accommodating them in the hospitals, where the spread of the disease is at a higher risk. The collected data including the test results and the contact information  of the patients in the medical vehicle are sent to the medical authorities. Then, the patients can have further treatment and advices upon the diagnosis in a remote fashion. This approach enables a rapid diagnosis while keeping the risk of crowd in the hospitals at minimum level.

\begin{figure*}[t]
\centering
{\includegraphics[draft=false,width=\textwidth]{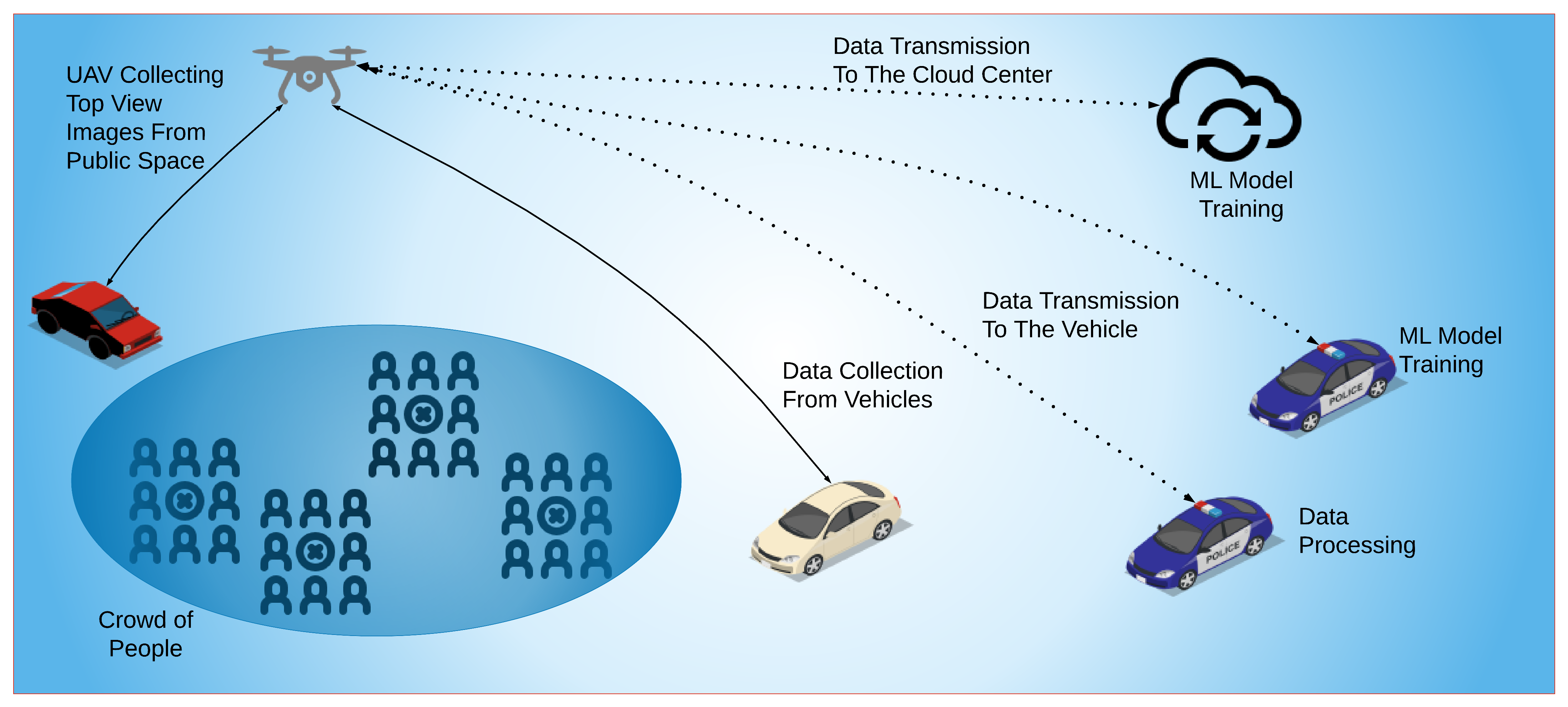} }
\caption{UAV-assisted social distance monitoring via ML, involving data collection from the vehicles, data processing in the vehicles and model training in the cloud center and vehicles.
}
\label{fig_UAV_socialDistanceMonitoring}
\end{figure*}

\subsubsection{Border Patrol}
In case of a worldwide pandemic, the travel of the people via public or private transportation may lead to long queues of vehicles before the checkpoints in the city entrances or even within the city depending on the spread of the disease, where the health status, follow-up of the interaction with infected individuals and travel allowances of the passengers are usually checked. In such a scenario, vehicular networking can be used to automatically transfer the passenger data from the vehicles to the RSUs or police cars on the side of the road, as illustrated in Figure~\ref{figBorderPatrol}. The passenger data can be collected by again reading the identity card of the passengers or their mobile phones from the RSUs via V2I communication.  Then, the data are transmitted from the RSUs to the cloud server for further investigation regarding the interaction of the passengers with infected individuals and  their travel allowances. Such a communication architecture results in the automatic monitoring of the passengers, stopping the vehicles only if there is any suspicion about the violation of regulations, saving time and fuel efficiency. Furthermore, this will minimize the human interactions in the monitoring process.

\subsubsection{UAV-assisted Crowd Monitoring}
During COVID-19, UAVs have been used to monitor the crowd of people for social distancing, by exploiting their ``bird's-eye-view'' and high flexibility~\cite{Amponis2021Nov}. \textcolor{black}{The UAVs offer several advantages such as minimizing the human interaction and reaching the inaccessible areas~\cite{Chamola2020May}.} \textcolor{black}{By combining the usage of UAVs and vehicles, drone-assisted vehicular networks (DAVNs) provide a ubiquitous connection for vehicles by efficiently integrating the communication and networking capabilities~\cite{Shi2018Jan}.} For instance, cameras mounted on the UAVs monitor the average social distance in the crowded areas in different cities, allowing the future predictions on the spread of the pandemic, for updating/improving the precautionary measures taken by local authorities. Here, ML is mostly used for the processing of the image data collected from the cameras of the UAVs to detect and track the individuals, as illustrated in Figure~\ref{fig_UAV_socialDistanceMonitoring}. These data are then sent to a cloud data server or a ground data station for model training since ML models require huge parallel processing resources, which cannot be conducted in the UAV hardware.	Once the ML model is trained, the UAVs are used over a crowded area for social distance monitoring.	In order to make more accurate predictions, UAVs can be used together with vehicles on the streets to benefit from the multiple sensors in the vehicles, such as cameras, proximity sensors and radars/lidars. Furthermore, the usage of vehicles in this scenario can provide additional computing hardware and power capability to improve the battery-limited performance of UAVs. Hence, the collected data from the UAV can be sent to a vehicle for data processing to deal with the battery constraints.

\begin{figure*}[t]
\centering
{\includegraphics[draft=false,width=\textwidth]{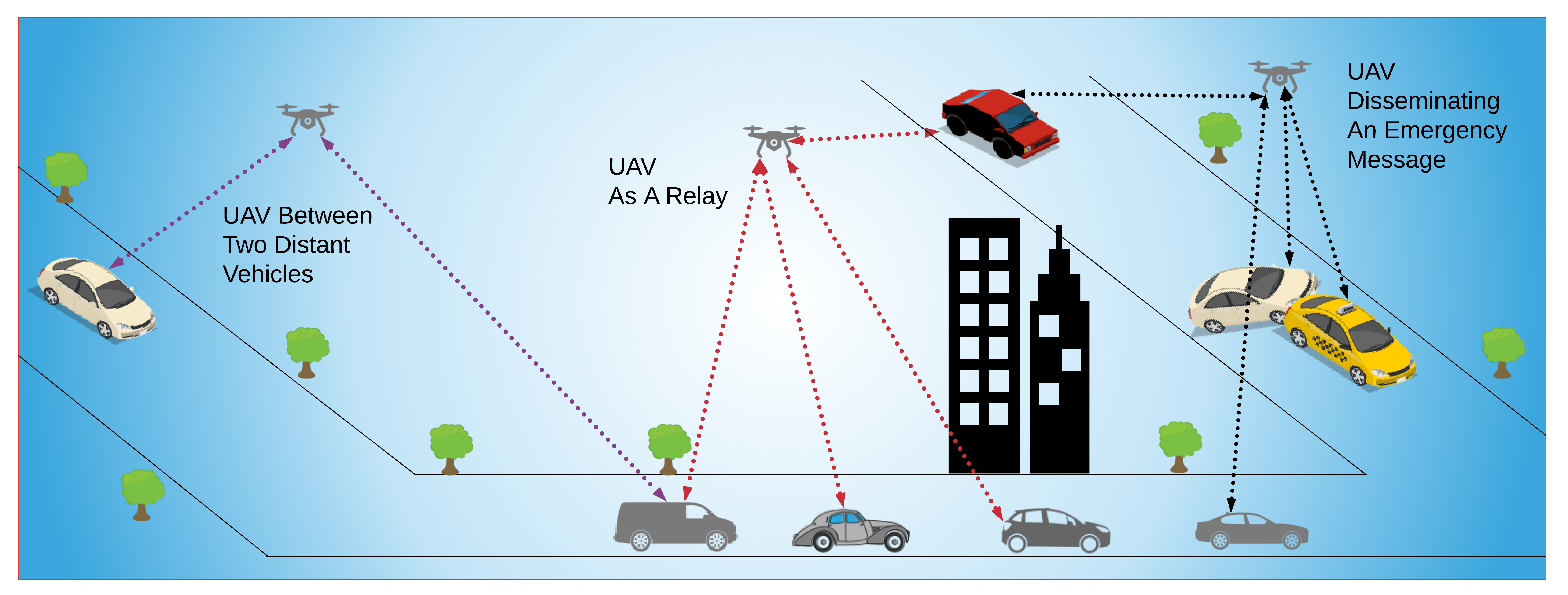} }
\caption{UAV-assisted vehicular communication. UAVs can act like message relays between ground vehicles, providing reliable communications in case of long distance, blockages between the vehicles and emergency.	}
\label{fig_UAV_assited_Vehicular}
\end{figure*}

When  the vehicular distribution in the cities is sparse in case of a pandemic, especially during a lockdown,	UAVs can be integrated into the V2V communication architecture to deliver information about the health status of passengers in the vehicles and path planning.  In this scenario, UAVs can act as message relays for the communication link between the vehicles, as illustrated in Figure~\ref{fig_UAV_assited_Vehicular}. For example, in	case of an accident, UAVs may quickly arrive at the incident scene and collect critical information from nearby vehicles.

{\color{black}In addition, in~\cite{uav_vehicular1}, UAVs are used  for emergency vehicle guidance  to plot the fastest path to intervene an accident by establishing a UAV-assisted vehicular communication. Similarly, UAV-assisted V2V scenario is also considered in~\cite{uavAssistedV2V} for the adaptive cruise control of vehicles in case of an emergency by integrating V2V with Thread, a secure wireless networking protocol based on IPv6 (Internet Protocol Version 6). 
These approaches should be extended to incorporate the additional challenges  of a pandemic, in which most of the hospitals serve at \textit{almost full} capacity. The transmitted data may then include the availability of the hospitals for accurate path planning. 
This approach can be adapted to a pandemic scenario, where the V2V communication carries the health information of the individuals in vehicles.
Furthermore, UAV-assisted vehicular communication is a part of intelligent transportation systems (ITSs) in smart city applications~\cite{uav_ITS,Shi2018Jan}.  }

\subsection{Requirements}

In the aforementioned applications, \textcolor{black}{the collected COVID-19 data have unique characteristics in terms of privacy, flexibility, coverage and prioritization, thus, demand novel vehicular network design approaches to take into account the highly mobile and heterogeneous architecture of vehicular networks, as detailed below.}

\subsubsection{Privacy}

\textcolor{black}{One of the main challenges in the above applications is preserving the privacy during both COVID-19 data collection and its transmission over the vehicular network~\cite{Abouelmehdi2017Jan}. Similar to the traditional medical data privacy, the COVID-19 data collected from a user should be available to only the user itself and the central authority, thus, personal data protection against large scale malicious activities must be provided. However, COVID-19 data should be processed and analyzed by government officials and companies to reduce the spread of the virus, which could curtails the privacy rights of patients. This concern is currently under discussion for regulatory issues.} Considering  the  privacy  during  transmission, especially in the cases of building a connection to a central entity via multi-hop V2V links, the privacy of the information must be preserved from the other vehicles in the link and the RSUs. 

\subsubsection{Flexibility}  

In conventional V2V networks, standard message structures include Cooperative Awareness Message (CAM), consisting of vehicle data, including location, heading and speed information, and Decentralized Environmental Notification Message (DENM), consisting of emergency notification data, including event cause code, sub-cause code and severity information, defined by ETSI~\cite{ETSI}. Since the data transmitted in CAM and DENM have fixed sizes, these message structures employ fixed frame size. Moreover, CAM and DENM enforce strict timing. CAM is sent periodically at least once every second, whereas DENM is an event-based notification message with different latency constraints for each event. However, the data specifically collected for pandemic prevention are highly unstructured and variable in both frame size and periodicity. The travel histories of different users may vary in size significantly. Moreover, the availability of sensor readings might be highly sporadic in the network. \textcolor{black}{The variable size and periodicity of the COVID-19 data requires the development of adaptive and flexible schemes.} For instance the resource allocation and management algorithms re-optimized for data with such unstructured features, while taking into account the dynamic network topology caused by the vehicle mobility.

\subsubsection{Coverage} 
\textcolor{black}{Different from the traditional medical data collection, the a wider coverage  should be required for the collection of the COVID-19 data. Hence, accurate prediction and documentation demand a fine coverage of the collected COVID-19 data.} The aforementioned vehicular applications are expected to improve the coverage of the collected data due to the common use of vehicles in daily life. However, it may not be always possible for a vehicle to transmit its collected pandemic data to the cloud server, due to the dis- and re-connections in the link caused by the dynamic nature of the vehicular channel. Furthermore, if the pandemic data are collected sparsely and dispersively, it is of a great challenge to transfer the collected pandemic data to the cloud data centers because of the lack of infrastructure to collect data from the vehicles, e.g., rare deployment of the RSUs in suburban areas. Then, it may be possible to have no information on pandemic status for some suburban regions, preventing the authorities from making quarantine decision for such regions. Therefore, efficient multi-hop data routing protocols are required for vehicular domain, aiming to increase the coverage of the network.   

\subsubsection{Prioritization} 
\textcolor{black}{Compared to traditional medical data, the COVID-19 data has higher priority for transmission and delivery to the authorities which can take necessary precautions, especially for the regions lacking the necessary infrastructure to collect data from vehicles.} In such regions, the data is cached by the vehicles and transmitted to the central authority as soon as they establish a connection to the infrastructure. In the cases of having limited access to the infrastructure, the vehicles need to choose which message to send. Here, a message prioritization criteria is required considering both the content and freshness of the message. Regarding the content, the data of the users tested positive for COVID-19 should have a higher priority than the others. Moreover, age of information (AoI), as a measure of freshness of the pandemic data, needs to be considered for the prioritization, since the outdated pandemic data may not be useful to the authorities in determining the pandemic spread of the region. Hence, pandemic data requires content-based prioritization mechanisms alongside with AoI-based caching protocols, mainly considering the locations where the vehicles have limited access to the infrastructure.

\section{Open Challenges and Future Research Directions}
In this section, we provide an extensive analysis on the research challenges in privacy, flexibility, coverage and prioritization, which are briefly discussed above, and provide related solutions as a future direction of research.


\subsection{Privacy Preserving Mechanisms}

\textcolor{black}{Privacy protection of the pandemic data must be handled during both data collection and its transmission over the vehicular network, taking into account that it consists of content-sensitive information as user name, ID and travel history~\cite{Sajid2016May}.} \textcolor{black}{In particular, the main function of the mobile tracing apps should include collecting information aiming at the general public, ability to alert on individuals at risk of infection, quarantine surveillance/monitoring for the confirmed infected individuals~\cite{Altmann2020Aug,BibEntry2021Mar}. Nevertheless, there is a trade-off between the privacy risk and the functionalities of the contact tracing apps. For instance, the privacy-preserving approach may not support any additional functionality for disease surveillance, which should be taken into account jointly~\cite{Wirth2020Nov}. Hence, the structure of the contact tracing app should be handled considering several issues such as use cases for different diseases; data transmission/collection technology, e.g., Bluetooth; and data-protection. The privacy preserving mechanisms during data collection can be achieved by developing decentralized data collection and storage~\cite{privacy_covid2}. In order to provide data-privacy, instead of collecting the data of all individuals in a cloud server, the pandemic data can be collected and cached in a decentralized manner, minimizing the vulnerability of the whole data being hacked.} To practically realize this approach in a vehicular network, each vehicle can act as the nodes of this decentralized architecture to form a cluster of the data of its own. Then, a device in the vehicle assigns a unique code to each of the passengers and share these data with each individual's cell phone without transmitting the information to the cloud data center. When an individual is subsequently diagnosed positive for COVID-19, he/she can input the diagnosis result into his/her cell phone while informing the other passengers in the cluster as well as the authorities. Thus, only the information belonging to the clusters containing at least one positively diagnosed person is shared with the authorities. This reduces the risk of large scale malicious activities during data transmission, by preventing the unnecessary share of the information of people in the  clusters without any positively diagnosed person.

Regarding the data transmission, vehicular networks have previously incorporated privacy preserving methods categorized as pseudonymization, randomization and aggregation~\cite{privacysurvey}. Pseudonymization is a procedure that alters the personal IDs in the data with the artificially created IDs. However, pseudonymization of the pandemic data is not an option since the ID and the name of the user is also desired to be transmitted to the central authority. On the other hand, randomization is a technique of perturbing a piece of the data before transmission, to hide the exact information, which leads to a decrease in the information accuracy. Similarly, aggregation techniques, targeting to preserve the privacy of the raw data, lead to decreased information accuracy. Even though aggregation methods can be applied to some of the pandemic applications like UAV-assisted monitoring for preserving data privacy, in general, they cannot be applicable to the applications where the raw data itself is required, such as border patrol applications. Hence, in general, the corruptions on the data due to the use of randomization and aggregation techniques may incur inaccurate results in data processing. As a result, deeper investigation is required to protect the privacy of the information, taking into account the aforementioned vehicular pandemic applications requiring accurate data with the personal information content.

\subsection{Resource Allocation}

Resource allocation is a challenge for pandemic applications since the data is unstructured and variable in both size and periodicity. In the literature, there exist some studies on adaptive protocol design targeting such data characteristics. In~\cite{flexibility_2}, the authors propose an adaptive bandwidth assignment mechanism for video transmission, by allocating more bandwidth to the high priority packets, which are the inter-frame packets, and less bandwidth to the low priority packets, which are the predicted frame packets of H.264 standard video coding mechanism, ensuring that every packet is sent at the minimum sending rate for fairness. In~\cite{flexibility_3_ITS}, an optimal rate-adaptive in-platoon data dissemination technique is proposed. The study approaches to the resource allocation problem as an end-to-end optimization problem for a multi-hop communication system, targeting to minimize the end-to-end latency, and investigating the impact of packet and platoon sizes on the latency. The usage of these techniques for the optimal resource allocation in vehicular pandemic applications requires addressing the flexibility and prioritization requirements through the formulation of a content-based prioritization mechanism. 

\subsection{Data Caching}

In addition to preserving privacy, data caching can be utilized to satisfy the flexibility, coverage and prioritization requirements of the pandemic data. Regarding the caching mechanisms in vehicular domain, unlike other wireless terminals, such as smartphones or sensors, vehicles do not have strict energy or memory constraints. Hence, in vehicular applications, the simplest caching approach is to cache all the data overheard over the wireless channel, which is called \emph{cache everything in the air}. However, indiscriminate caching can waste network resources and reduce the caching efficiency~\cite{ondemandCaching}. Therefore, connectivity-aware caching algorithms are investigated with the goal of maximizing the total caching utility, which is introduced as a function of caching gain and caching cost. Additionally, the effect of cache size on the delivery rate and latency has been analyzed for different caching mechanisms. Moreover, for the video streaming applications, bit-rate adaptive and scalable caching mechanisms are proposed, considering the quality of experience as a performance metric. \textcolor{black}{Furthermore, AoI-based caching mechanisms~\cite{AoI,caching_AoI}
running at RSUs are investigated with the objective of maximizing the freshness of the vehicle received information. These studies need to be extended for the pandemic applications in vehicular networks, with the goal of increasing the coverage, when the access to the infrastructure is not available, as well as providing more efficient and flexible resource allocation, and reduced access delay.} The caching utility and network performance of an adaptive caching algorithm need to be investigated by incorporating a prioritization mechanism for accurate and fast data extraction while satisfying a fine coverage of the variable size pandemic data. Moreover, the AoI-based caching mechanisms for uplink vehicular channel need to be studied, while taking account of the V2V link durability for the pandemic applications.

\subsection{Vehicular Data Routing}

\textcolor{black}{The coverage of the vehicular network for the pandemic applications can be extended by employing flexible routing mechanisms~\cite{RoutingPriority}. There are numerous studies related to multi-hop routing protocols in vehicular ad hoc networks (VANETs)~\cite{Ghori2018May}.} Flexible position-based routing mechanisms have been proposed with the goal of offering session continuity and reachability to the infrastructure, by dynamic V2V and V2I link selection based on the real-time feasibility of the links\cite{baldessari2006}.
Additionally, for the transmission of the health information of a patient in an ambulance to the hospital, a routing protocol targeting the V2V links with smaller link breakage probability is investigated in \cite{vehiHealth}. 
Moreover, privacy preserving opportunistic routing protocols for vehicular networks utilizing Paillier homomorphic encryption scheme is employed to enable the vehicles to compute and compare the similarities in their routing metrics without compromising the data privacy in~\cite{ePRIVO}.
Furthermore, priority and broadcast based routing protocols in VANETs consider the different quality of service requirements for the packets at different priority classes \cite{RoutingPriority}.
For the usage of these routing algorithms in pandemic applications, these routing algorithms need to be extended by considering the privacy and prioritization  requirements of the pandemic data.


\subsection{Machine Learning Solutions}


\textcolor{black}{ML-based solutions can be employed for data aggregation to help flexible resource allocation mechanisms by reducing the data transmission cost in the network. Decentralized ML approaches, such as federated learning (FL)~\cite{Rieke2020Sep}, have been recently proposed for vehicular communication to reduce the data processing and inference complexity~\cite{elbir2020federated}.} In FL, the training of the ML model is performed without the transmission of the whole dataset, instead, only the model updates, i.e., the gradients of the model parameters, are sent to a cloud server for model training, preserving the data privacy. \textcolor{black}{FL can be used for in-vehicle infection diagnosis application, where  the input of the learning model can be the collected data from the individuals, such as infection test data and CT (computerized tomography) images, and the output can be the classification of whether the subject is COVID-19 positive/negative~\cite{Zhang2021Feb,ML_COVID}. For instance, a fusion-based FL approach is proposed in~\cite{Zhang2021Feb} or medical diagnostic image analysis to detect COVID-19 infections. This approach involves training a learning model on local datasets and schedule the model-based fusion of the clients' training time, thereby providing higher learning accuracy, communication-efficiency and fault tolerance.} FL can also be helpful in multi-UAV-assisted crowd monitoring applications, in which the input of the learning model can be the image data obtained from the UAV cameras and the output can be the risk factor in accordance with the crowdedness. {\color{black}FL-based intrusion detection is another application that can be helpful for healthcare systems. In~\cite{Siniosoglou2021Jun}, an FL architecture for medical cyber-physical systems (MCPS) is proposed to induce enhanced security for model training. In particular, a generative adversarial network (GAN) is employed in a multi-layer FL scheme, wherein the anomalies indicating cyber-attacks to the medical network are detected with improved accuracy.} In both VANET and multi-UAV scenario, the convergence rate of FL greatly depends on the quality of the communication link between the vehicles and the cloud server. However, the availability of the network resources vary significantly among the nodes due to both sparse spatial distribution and heterogeneous architecture of the vehicular networks. This makes FL-based model training a challenging task for vehicular networks, wherein the device scheduling and resource management during model training should be explicitly optimized with the objective of maximizing the FL convergence rate.  
%
%
%

\section{Summary and Future Outlook}
This article  focuses  on the usage of vehicular networking in combating a pandemic. We investigate vehicular applications in public transportation, in-vehicle diagnosis, border patrol and UAV-assisted social monitoring. Each of these applications is demonstrated to introduce new challenges in terms of privacy, flexibility, coverage and prioritization of data. We believe that this work provides a comprehensive insight for research and product development in the usage of vehicular networking for reducing the spread of the pandemic. {\color{black}In particular, we provide a comprehensive analysis of the requirements and corresponding research challenges, and summarize the following potential research gaps to be studied for future work:

The privacy is the most crucial part for data collection and processing in VANET-based tracing applications. Apart from protecting the data, the implementation of the privacy-tracing apps should taken into account the several aspects, such as the design of the data transmission technologies and flexibility of the usage for different diseases. In addition, new approaches including ML/FL-based techniques can offer new opportunities for low-complexity, yet-accurate solutions. 

Resource allocation is critical to manage the collected data for efficient processing. In particular, low-latency solutions are required to be implemented for VANET applications in the next generation communication systems. In this respect, rather than allocating the resource in terms of physical layer data performance metrics, e.g., wireless channel quality, content-based approaches can provide more efficient and effective solutions to prioritize the resource management process.

Due to mobility nature of VANETs, data caching is an important aspect of the vehicular-based disease detection and tracing operations for effective use of memory. In this respect, efficient caching schemes, such as AoI-based caching techniques have potential  to achieving low-complexity and low-latency for future research.

ML-based solutions have recently been interesting for several model-based problems. In VANET-based applications, ML can provide extracting the knowledge inheriting in the data to lower the computational complexity involved during data transmission in VANETs. In addition, FL-based schemes may be integrated to UAV or vehicular scenarios to exploit the distributed nature of UAV and vehicular networks.

}

\section*{Acknowledgment}
This work was supported in part by Ford Otosan, CHIST-ERA grant CHIST-ERA-18-SDCDN-001 and the Scientific and Technological Council of Turkey 119E350.

	\bibliographystyle{elsarticle-num}
	\bibliography{references_overview1}

	\balance

\end{document}